\documentclass[%
 reprint,
 amsmath,amssymb,
 aps,
 prx
]{revtex4-2}

\usepackage{graphicx}% Include figure files
\usepackage{dcolumn}% Align table columns on decimal point
\usepackage{bm}% bold math
\usepackage{xcolor}
\usepackage{dsfont}
\usepackage{amsmath}
\usepackage{amssymb}
\usepackage{subcaption}
\usepackage[colorlinks=true, linkcolor=blue,citecolor=blue]{hyperref}
\newcommand\fracd[2]{\frac{d {#1}}{d {#2}}}

\newcommand\cala{\mathcal A}

\newcommand\eye{\tensor 1}

\newcommand\uvec[1]{\hat {#1}}

\newsavebox{\mySaveBoxMath} % M for math
\newsavebox{\mySaveBoxText} % T for text

\renewcommand{\vec}[1]{\boldsymbol{#1}}
\renewcommand{\uvec}[1]{\hat{\boldsymbol{#1}}}
\renewcommand{\tensor}[1]{\underline{\boldsymbol{#1}}}
\newcommand\utensor[1]{\tensor{\hat {#1}}}
\newcommand\tensop[1]{\widehat{\tensor{#1}}}

\newcommand{\compactcross}[1]{\hspace[-0.1em]\times\hspace[-0.1em]}

\renewcommand{\tensop}[1]{\widetilde{\tensor{#1}}}
\newcommand{\ham}{\mathcal H}

\begin{document}

\preprint{APS/123-QED}

% \title{Spin-orbit interactions in non-uniformly moving media}
\title{Spin-orbit interactions induced by light drag in moving media}

\author{Aymeric Braud}
 \email{aymeric.braud@laplace.univ-tlse.fr}
\author{Renaud Gueroult}
 \affiliation{LAPLACE, Université de Toulouse, CNRS, INPT, UPS, 31062 Toulouse, France}

\date{\today}% It is always \today, today,

\begin{abstract}
Spin–orbit interactions (SOIs) of light are manifestations of coupling between components of light's angular momentum. They are at play in most basic optical processes, offering opportunities both to understand their fundamental origin and to control light in novel ways. Because SOIs become significant at subwavelength scale, they have largely been explored in the context of inhomogeneous materials exhibiting wavelength-scale structures, and notably metamaterials. Here we demonstrate that spin-orbit interactions can in fact analogously emerge in moving matter through the well-known light-dragging effects. SOIs in moving media are shown to manifest through a Berry phase induced by vorticity, which then leads to a rotation of the wave's polarization. In bringing together electrodynamics of moving media and SOIs of light, our work not only paves the way for the discovery of new fundamental effects but also uncovers novel means to harness SOIs to control light.
\end{abstract}

\maketitle

%\tableofcontents

%\section{Introduction}\label{sec:intro}

\section{Introduction}

Electromagnetic waves are described by vector fields. They thus possess, in addition to scalar wave's intensity and phase spatial distributions, additional degrees of freedom in the form of polarization. The latter, which is associated with the photon's spin, corresponds to the spin angular momentum of light~\cite{Beth1936}, while the spatial degrees of freedom determine the orbital angular momentum of light~\cite{Poynting1909,Allen1992}.

Owing to the vectorial nature of Maxwell's equations, polarization is, however, not independent of the wave's spatial structure: one influences the other and reciprocally~\cite{liberman_spin-orbit_1992,Allen1996,Allen1999}. Fundamentally, this coupling between light's spin and orbital degrees of freedom is a manifestation of spin-orbit interactions (SOIs)~\cite{bliokh_spinorbit_2015}, akin to SOIs of quantum particles~\cite{Mathur1991} or of electrons in solids~\cite{Berard2006}. An attractive and singular aspect of light SOIs is that while weak in large-scale classical optics, they become salient at sub-wavelength scales. They notably manifest in the wavelength-scale light-matter interactions enabled by nano-optics, nano-photonics and plasmonics, which has led to numerous developments harnessing SOIs to structure light in different degrees of freedom~\cite{Cardano2015,Xiao2016,Ling2017,Shi2022,Sheng2023,DiColandrea2023}. 
%The finding of these basic effects, combined with the development of metamaterials, has then sparked numerous developments harnessing SOIs to structure light in different degrees of freedom~\cite{Cardano2015,Xiao2016,Ling2017,Shi2022,Sheng2023,DiColandrea2023}.
A large part of these developments revolved more particularly around two remarkable observable effects: the Berry phase~\cite{berry_quantal_1984,Berry1987,Bliokh2008} and the spin Hall effect~\cite{Onoda2004} (also known as optical Magnus effect~\cite{bliokh_modified_2004}). 

The Berry phase~\cite{berry_quantal_1984,Berry1987} describes variations of polarization along the light ray. It stems from the non-trivial evolution of either the polarization state or the wavevector in its corresponding parameter space, and adds up to the dynamical phase. An archetypal example is the Rytov rotation~\cite{rytov_sur_1938,Vladimirskiy1941}, where an extra phase, known as the spin-redirection or Rytov-Vladimirsky-Berry phase, arises from a rotation of the polarization plane due to variations in the direction of propagation. This effect typically occurs when the wavevector is redirected by inhomogeneities in an isotropic media~\cite{bliokh_geometrodynamics_2009}, or by a helically wound optical fiber~\cite{Ross1984,Tomita1986}, the key ingredient being the non-planarity of the light ray. Note that while the Rytov's rotation is often studied in relation to its associated effect on the ray through the spin-Hall effect~\cite{Sadykov1994}, it has also been recently suggested that it could contribute significantly to light depolarization in astrophysics, where ray bending occurs due to scattering in a randomly inhomogeneous turbulent plasma~\cite{Zhang2024}.

\begin{figure*}
    \centering
    % \begin{subfigure}{33mm}
    %     % \includegraphics[width=\linewidth]{index_refraction.pdf}
    %     % \includegraphics[width=\linewidth]{transverse_drag.pdf}
    %     % \includegraphics[width=\linewidth]{refraction_index_gradient.pdf}
    %     % \vfill
    %     % \includegraphics[width=\linewidth]{refraction_velocity_gradient.pdf}
    %     % \includegraphics[width=\linewidth]{refraction_index_and_velocity_gradientV2.pdf}
    %     \includegraphics[width=\linewidth]{refraction_index_gradient_with_zoom.pdf}
    %     \includegraphics[width=\linewidth]{refraction_velocity_gradient_with_zoom.pdf}
    % \end{subfigure}
    % \includegraphics[width=0.2\linewidth]{ray_notations.pdf}
    % \includegraphics[width=16.5cm]{Fig.pdf}
    % \begin{subfigure}{0.8\linewidth}
    %     % \includegraphics[width=\linewidth]{ray_in_velocity_field_with_ray_notations.pdf}
    %     \vspace{-0.65cm}
    % \end{subfigure}
    \includegraphics[width=\linewidth]{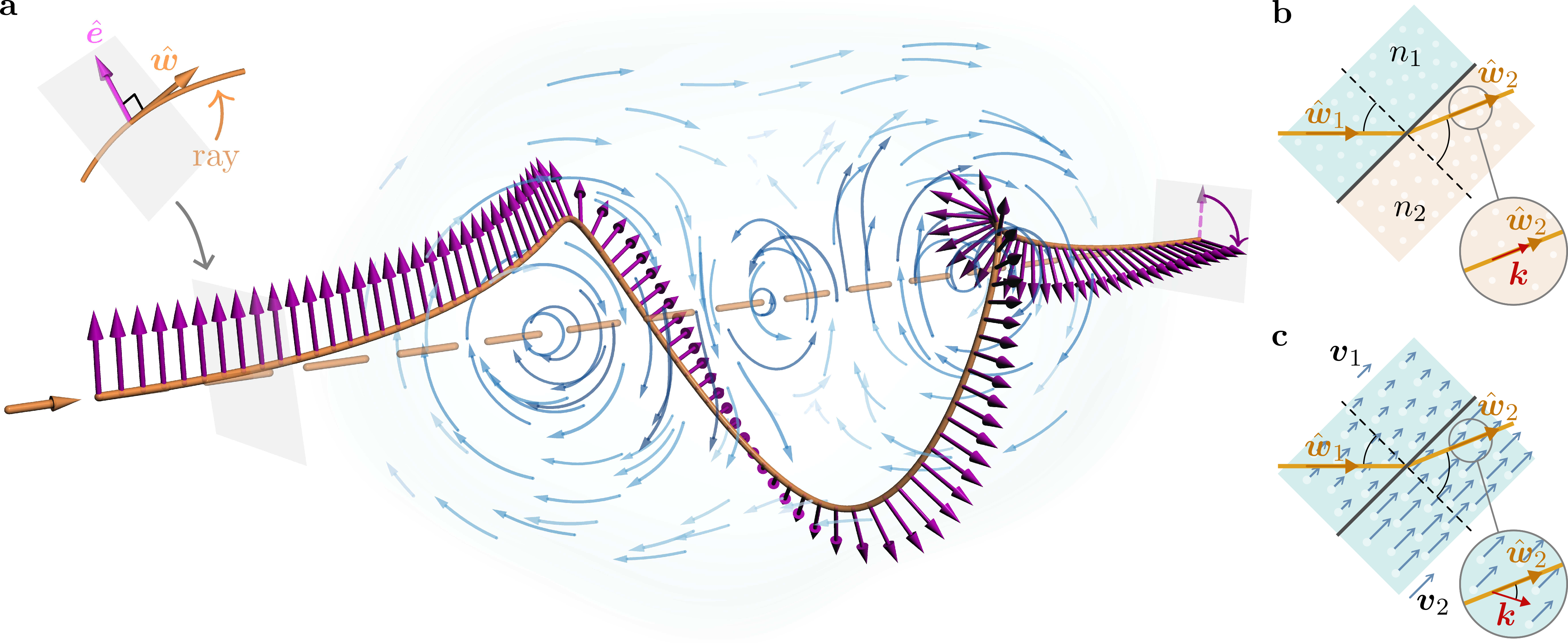}
    \caption{\textbf{a} Sketch of the non-planar ray trajectory (orange) induced by light drag in a non-uniform moving medium (blue streamlines), and of the associated rotation of the wave polarization $\uvec e$ (magenta) that then results from the induced spin-orbit coupling. The insets show how a change in ray direction $\uvec w$, as required to create a non-planar ray, can equally emerge from: \textbf{b} a change in refraction index, \textbf{c} wave drag in a moving medium.}
    %\caption{Illustration of a non-planar ray trajectory (orange) induced by light drag in a non-uniform moving medium (blue streamlines), and of the associated rotation of the wave polarization $\uvec e$ (magenta) that then results from the induced spin-orbit coupling controlled by the local vorticity. The insets show how a change in ray direction $\uvec w$, as required to create a non-planar ray, can equally emerge from a) a change in refraction index and b) wave drag in a moving medium.}
    %\caption{Change in ray direction $\vec w$ due to a) a change in refraction index and to b) a uniform motion (wave drag), as seen in the lab-frame $\Sigma$ and in the rest-frame $\Sigma'$; c) non-planar ray trajectory induced by a non-uniform motion, which then leads through spin-orbit coupling controlled by the local vorticity to a rotation of the polarization. \textcolor{blue}{To illustrate what I had in mind, needs work.}}
    \label{fig:general_motion}
\end{figure*}

While, as classically considered for Rytov's rotation~\cite{bliokh_geometrodynamics_2009}, a variation in the direction of propagation can stem from a variation in refractive index (Fig.~\ref{fig:general_motion}.\textbf{b}),  it has long been established that it can also be induced by motion in the form of light-dragging effects (Fig.~\ref{fig:general_motion}.\textbf{c}). In its simplest form, a wave normally incident on a transversely moving isotropic dielectric experiences, for an observer at rest, a change in its group velocity direction~\cite{player_dispersion_1975,jones_fresnel_1972,jones_aether_1975}. Recognizing its similarities with the longitudinal drag studied by Fresnel~\cite{fresnel_lettre_1818}, this effect is known as Fresnel's transverse drag. Now, while a uniform velocity only leads to a deflection and cannot yield a non-planar ray, it has been demonstrated that a non-uniform velocity field can bend light rays propagating through it~\cite{leonhardt_optics_1999,rozanov_firstorder_2006}. A non-uniform velocity field can then, similarly to an inhomogeneous refractive index, lead to non-planar rays, which should be the source of spin-orbit interactions manifesting as polarization rotation as illustrated in Fig.~\ref{fig:general_motion}.\textbf{a}.

In this work, we demonstrate that spin-orbit interactions emerge even in a homogeneous isotropic medium if this medium is in non-uniform motion, manifesting notably in new contributions to the evolution of the wave's polarization. Specifically, we show that velocity non-uniformity is the source of a geometric Berry phase induced by the non-planar trajectory, but also of a supplemental phase stemming directly from the local vorticity. Both effects contribute to the wave's phase evolution, leading to an overall polarization rotation, and associated corrections to the ray trajectory. We then derive formula for these effects in the canonical example of a rotating cylinder, showing in passing that our results reconcile Thomson and Fermi's prediction of polarization drag in a rotating dielectric~\cite{Thomson1885,Fermi1923} with the helical trajectory induced by light drag~\cite{padgett_polarization_2006,Franke-Arnold2011}. In demonstrating that spin-orbit coupling of light can be elicited through motion, our results not only make it possible to leverage moving media to develop novel means to manipulate light's degrees of freedom, but may also find useful analogs for quantum particles~\cite{Dong2021,Din2024}.

\section{Results}

\subsection{Non-planar trajectory}
Let us first illustrate how a non-uniform motion can lead to non-planar rays. As first shown by Minkoswki~\cite{Minkowski1908}, motion is the source of a magnetoelectric coupling: a medium that is isotropic at rest exhibits bi-anisotropic constitutive relations when seen in motion~\cite{Kong2000}. These anisotropic relations, coupled with Maxwell's equations, then describe wave propagation in a moving medium as seen by an observer at rest in the laboratory. For a isotropic non-dispersive medium with a uniform rest-frame refractive index $n'$ and a velocity field $\vec\beta(\vec x)=\vec v(\vec x)/c$ with $c$ the speed of light, we find as derived in Appendix~\ref{app:A} that the electric field $\vec E$ verifies to first order in $\beta\ll 1$ the wave equation
%\begin{multline}
%\label{Eq:wave_eq}
%    \tensop D =-\vec\nabla\times\vec\nabla\times~-n'^2\partial_{ct}^2~ 
%    \\+(n'^2-1)\left[\vec\beta\times\vec\nabla\times\partial_{ct}~+\vec\nabla\times\vec\beta\times\partial_{ct}~\right]
%\end{mutline}
\begin{multline}
\label{Eq:wave_eq}
\vec\nabla\times\vec\nabla\times\vec E+n'^2\partial_{ct}^2\vec E= 
    \\(n'^2-1)\left[\vec\beta\times\vec\nabla\times\partial_{ct}\vec E+\vec\nabla\times\left(\vec\beta\times\partial_{ct}\vec E\right)\right].
\end{multline}
The first line in Eq.~\eqref{Eq:wave_eq} is the standard wave equation for this medium at rest, whereas the second line captures corrections from motion. Looking for plane waves solutions $e^{i(\omega t-\vec k\cdot\vec x)}$, we first neglect the derivatives of the velocity field in Eq.~\eqref{Eq:wave_eq} and, as shown in Appendix~\ref{app:B}, obtain two degenerate transverse modes verifying the same dispersion relation $\ham_0=0$ with the dispersion function
%\begin{equation}
%    \ham_0(\vec x,\omega,\vec k)=\frac{\omega^2}{c^2}-k^2+(n'^2-1)\frac{\omega}{c}\left[\frac{\omega}{c}-2\vec k\cdot\vec\beta(\vec x)\right]
%\end{equation}
\begin{equation}
\label{Eq:hamilton_0}
    \ham_0(\vec x,\omega,\vec k)=n'^2\frac{\omega^2}{c^2}-k^2-2(n'^2-1)\frac{\omega}{c}\vec k\cdot\vec\beta(\vec x).
\end{equation}
Here we will work with right- and left-circularly polarized eigenmodes, denoted respectively by $+$ and $-$ subscripts. Eq.~\eqref{Eq:hamilton_0} is
consistent both with Leonhardt's~\cite{leonhardt_optics_1999}, and with the first order limit of the dispersion relation obtained from Gordon's optical metric~\cite{gordon_zur_1923}.

Considering the dispersion function $\ham_0$ as the Hamiltonian of a point-particle~\cite{littlejohn_geometric_1991}, Hamilton's equations then give the ray equations of geometrical optics, which describe the evolution of light rays in phase space $(\vec x, \vec k)$~\cite{tracy_ray_2014}. In particular, as shown in Appendix~\ref{app:C}, the dependence of $\ham_0$ on $\vec k\cdot\vec\beta$ manifests in the ray equation for the position $\vec x$ as a velocity dependent drag~\cite{player_dispersion_1975}, that is a finite angle between the ray direction $\uvec w$ (i.e. the group velocity direction) and the wavevector $\vec k$ due to motion. This deviation, shown in Fig.1.\textbf{c}, is characteristic of motion-induced anisotropy~\cite{Braud2025}. The ray equations for the ray position $\vec x$ and the wavevector $\vec k$ can also be combined to give
\begin{equation}\label{eq:ray_equation}
    \fracd{^2\vec x}{s^2}=\mathcal{N}\bm{\nu}\times\fracd{\vec x}{s}.
\end{equation}
Here $s$ is a ray parameter taken as the ray length so that $\uvec w=d\vec x/ds$, $\mathcal{N}=n'(1-1/n'^2)$ where $1-1/n'^2$ is Fresnel's dragging coefficient, $\vec\nu=\vec\nabla\times\vec\beta$ is the local vorticity (normalized by $c$), and overhats denote unit vectors. We recognize here, as already noted by Leonhardt~\cite{leonhardt_optics_1999}, that the dynamics of light rays is analogous to that of a charged particle in a magnetic field, with the vorticity $\bm{\nu}$ playing the role of the magnetic field. This illustrates that a non-uniform velocity field with non-zero vorticity will lead to a non-planar ray, that is to a ray with non-zero torsion.

\subsection{Motion induced spin-orbit coupling}

To confirm the induced effect of motion on spin-orbit coupling, we now consider as usual next-order corrections to classical geometrical optics~\cite{rytov_sur_1938}. Specifically, we search for mode solutions of the form of quasi-plane waves $a_\pm\uvec \eta_\pm e^{i\Phi}$~\cite{sluijter_general_2008}. These waves are locally plane, with an envelope $a_\pm\uvec \eta_\pm$ that is no longer constant but still varies slowly compared to the dynamical phase $\Phi$. Here $\uvec \eta_\pm$ are the mode unit polarization eigenvectors and $a_\pm=|a_\pm|e^{i\psi_\pm}$ is the complex amplitude which captures the extra phase $\psi_\pm$ that adds small corrections to $\Phi$. Typically, quasi-plane waves are sought to model propagation in inhomogeneous media~\cite{Kravtsov2011}, assuming that the wavelength $\lambda$ is small compared to the inhomogeneity lengthscale. A key contribution of our work is to extend this idea to a non-uniform velocity $\vec v(\vec x)$, and we accordingly require here $\varepsilon\doteq\lambda\|\vec\nabla\vec\beta\|_\infty/\beta$ to be small.

Formally, using this quasi-plane wave ansatz in the wave equation and expanding to first order in $\varepsilon$ yields the Hamiltonian matrix $\tensor{ \ham} = \ham_0\tensor{\mathds{1}}-\tensor{ \ham_1}$ describing waves to first order in $\varepsilon$. As presented in Appendix~\ref{app:D}, we find that the circularly polarized modes are independent with scalar first-order corrections $\ham_{1\pm}$. In static inhomogeneous media these first-order corrections stem entirely from spin-orbit interactions, which integrated along the ray classically give the Berry phase of each mode~\cite{bliokh_geometrodynamics_2009}. For our moving medium, we demonstrate importantly that these corrections arise from two distinct contributions $\ham_{SOI\pm}$ and $\ham_{GC\pm}$, each contributing to the slow phase evolution through $d\psi_{\pm}/ds=\ham_{SOI\pm}+\ham_{GC\pm}$.

The first one is the spin-orbit contribution
\begin{equation}\label{eq:H_SOI}
    \ham_{SOI\pm}=-\dot {\vec x}\cdot\vec \cala^{(x)}_\pm-\dot{\vec k}\cdot\vec \cala^{(k)}_\pm
\end{equation}
with $\vec\cala^{(x)}_\pm$ and $\vec\cala^{(k)}_\pm$ the Berry connections associated with respectively the ray position $\vec x$ and the wavevector $\vec k$, and a dot denoting a derivative with respect to the ray parameter $s$. This term is, analogously to SOIs in isotropic media~\cite{Bliokh2008}, the source of a geometrical Berry phase. However, this spin-orbit Hamiltonian now involves both $\vec\cala^{(x)}_\pm$ and $\vec\cala^{(k)}_\pm$ (as opposed to $\vec\cala^{(k)}_\pm$ only~\cite{Bliokh2008}) since the eigenvectors $\uvec\eta_{\pm}$ now depend both on $\vec x$ and $\vec k$. Yet, because the eigenvectors here are only functions of the ray tangent vector $\uvec w$, Eq.~\eqref{eq:H_SOI} can actually be recast in the compact form $\ham_{SOI\pm}=-\dot{\uvec w}\cdot\vec \cala_{\pm}^{(\hat w)}$, where we introduced $(\cala_{\pm}^{(\hat w)})_j=-i\uvec\eta_{\pm}^\dag\partial{\uvec\eta_{\pm}}/\partial{\hat w_j}$ the Berry connection associated with $\uvec w$. Computing it explicitly from Eq.~\eqref{Eq:wave_eq}, we find as derived in Appendix~\ref{app:D}
\begin{equation}\label{Eq:Connection_w}
    \vec\cala^{(\hat w)}_\pm=\pm\frac{\uvec w\cdot\uvec u}{1-(\uvec w\cdot\uvec u)^2}(\uvec w\times\uvec u)
\end{equation}
with $\uvec u$ a constant gauge vector, whereas, from Eq.~\eqref{eq:ray_equation}, $\dot{\uvec w} = \mathcal{N}\bm{\nu}\times\uvec w$. We thus show here unequivocally that $\ham_{SOI\pm}$ is non-zero even if we assumed a homogeneous isotropic medium. This result demonstrates that spin-orbit coupling can arise from a velocity non-uniformity, and more specifically from a finite flow vorticity $\nu$.

Writing $\ham_{SOI\pm}$ through $\vec\cala^{(\hat w)}_\pm$ further exposes that spin-orbit coupling arises from changes in the direction of $\uvec w$, which are here due to motion, just like it usually arises from changes in the direction of $\vec k$ in inhomogeneous isotropic media~\cite{bliokh_geometrodynamics_2009}. It in turn supports the interpretation of the motion induced SOI as a spin-redirection effect for non-planar ray trajectories. Indeed, recalling that due to the anisotropy induced by motion the eigenvectors $\uvec\eta$ are in the plane normal to $\uvec w$ (as opposed to the plane normal to $\vec k$), changes in the direction of $\uvec w$ are the source of rotations of the plane of polarization. Confirming this finding, a closer examination shows that Eq.~\eqref{Eq:Connection_w} is identical to $\vec\cala^{(k)}_\pm$ known in static inhomogeneous media~\cite{Bliokh2008}, other than for the dependence on $\uvec w$ instead of $\uvec k$, and that the difference is in fact captured in the velocity dependent $\vec\cala^{(x)}_\pm$.

Looking now at the second contribution to $\ham_{1\pm}$ complementing SOIs, we find that
\begin{equation}\label{eq:H_GC}
    \ham_{GC\pm}=\mp\frac12\mathcal{N}\uvec w\cdot\bm{\nu}.
\end{equation}
We refer to this term as gradient correction (GC), as it captures the effect of spatial variations of the velocity field. Interestingly, this term bears close similarities with corrections due to spatial variations in the Coriolis parameter in the ray Hamiltonian describing equatorial shallow water waves in geophysics~\cite{onuki_quasi-local_2020,perez_manifestation_2021,venaille_ray_2023}, which have analogously been shown to contribute, together with SOIs, to the dynamics of the wave. We also note that additional terms supplementing SOIs have already been reported in weakly anisotropic media~\cite{Bliokh2007,Ma2016}, though the latter differ in that they depend on the medium's properties directly, rather than on their spatial derivatives as is the case of $\ham_{GC\pm}$.

%\textcolor{orange}{Such a term depending on spatial derivatives of the medium's properties is similar to the one found in studies on geophysical waves~\cite{onuki_quasi-local_2020,perez_manifestation_2021,venaille_ray_2023} where it contributes along with SOI to modify the dynamics of the wave. Additional terms supplementing SOI are, moreover, reported in weakly anisotropic media~\cite{Bliokh2007,Ma2016} but these latter differ from the present $\ham_{GC}$ in that they depends on the medium's properties directly and not on their derivatives.} Although this additional term supplementing SOI is akin to that already reported in weakly anisotropic media~\cite{Bliokh2007,Ma2016}, it differs in that $\ham_{GC}$ depends on the spatial derivative of the velocity field. This is in contrast to anisotropic media where the first order Hamiltonian only depends on the medium's properties, and not on their derivatives.\textcolor{blue}{[Something on Lyon, Onuki who on the other hand have something similar. Also Emerich?]} 

Finally, while we will consider next the effect on polarization, we note that there will also be mode specific first-order corrections to the ray trajectory associated to each of these contributions to the slow phase, that is a spin-Hall effect~\cite{Bliokh2008}. A non-uniform velocity field will hence separate modes spatially through $\ham_{SOI}$ and $\ham_{GC}$, similarly to how an inhomogeneous Coriolis parameter affects the trajectory of inertia-gravity wave packets~\cite{perez_manifestation_2021,venaille_ray_2023}.

\subsection{Resulting effects on polarization}

\begin{figure}
    \centering
    \includegraphics[width=\linewidth]{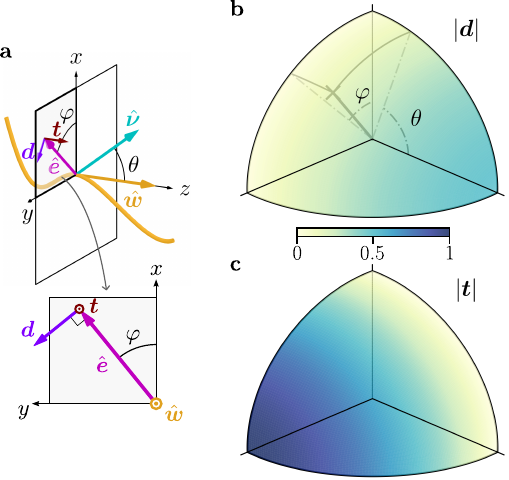}
    \caption{\textbf{a} Directions of the polarization drag $\vec d$ and parallel transport $\vec t$ contributions to the evolution of the polarization unit vector $\uvec e$, and \textbf{b}, \textbf{c} amplitudes of these two contributions in ($\theta,\varphi$) space, normalized by $\mathcal{N}\nu$.}
    \label{fig:TPvsGC}
\end{figure}

We showed that both SOI and gradient corrections are the source of a difference in the extra phase $\psi_{\pm}$ of the two modes. Each of these contributions, together with variations in the eigenvectors $\uvec\eta_{\pm}$ along the ray, is the source of a rotation of the polarization of the wave field $\vec E=\vec E_++\vec E_-$ obtained as the superposition of the two modes. Specifically, we show in Appendix~\ref{app:E} that the evolution of the unit polarization vector $\uvec e=\vec E e^{-i\Phi}/|\vec E|$ along the ray in our moving medium is governed by
\begin{equation}\label{eq:pola_evol_t+d}
\frac{d\uvec e}{ds} = \vec t+\vec d
\end{equation}
with $\vec t$ and $\vec d$ two orthogonal vectors in the plane normal to $\uvec e$ as illustrated in Fig.~\ref{fig:TPvsGC}.\textbf{a}.

%To see this we consider the unit polarization vector $\uvec e=\vec E/|\vec E|$, whose evolution along the ray verifies $d\uvec e/ds=t\uvec w+\vec d$ (see Methods).

The first contribution
\begin{equation}\label{eq:t}
\vec t =\mathcal N\left[\vec\nu-(\uvec w\cdot\vec\nu)\uvec w\right]\times\uvec e\
 \end{equation}
stems from the combined effects of $\ham_{SOI}$ and of variations of $\uvec\eta_{\pm}$, which together produce, as expected, a gauge independent effect. Remarking that Eq.~\eqref{eq:t} can be rewritten $\vec t=-(\uvec e\cdot\dot{\uvec w})\uvec w$, we find that it is precisely the variation expected from parallel transport~\cite{bliokh_geometrodynamics_2009}. This suggests that the slow phase accumulated due to $\ham_{SOI}$ is associated to the parallel transport of the physical polarization along the ray, similarly to the spin-redirection phase in inhomogeneous media~\cite{Haldane1986,Berry1987}. This result confirms that, in inducing non-planar rays, a non-uniform motion leads to a polarization precession akin to Rytov's rotation~\cite{rytov_sur_1938}.

%The first contribution 
%\textcolor{gray}{$\vec t=-(\uvec e\cdot\dot{\uvec w})\uvec w$ arises} \textcolor{orange}{$\vec t$ arising} from the combined effects of $\ham_{SOI}$ and of variations of $\uvec\eta_{\pm}$, which together produce, as expected, a gauge independent effect \textcolor{orange}{proves to be equal to $-(\uvec e\cdot\dot{\uvec w})\uvec w$. This is precisely the variation imposed by the parallel transport of a transverse vector along a curve~\cite{bliokh_geometrodynamics_2009}, aligned with the ray direction as shown on Fig.~\ref{fig:TPvsGC}(a). This suggests} \textcolor{gray}{It is aligned with the ray direction, suggesting} that the slow phase accumulated due to $\ham_{SOI}$ is associated to the parallel transport of the physical polarization along the ray, similarly to the spin redirection phase in inhomogeneous media~\cite{Haldane1986,Berry1987}. This result confirms that, in inducing non-planar rays, a non-uniform motion leads to a polarization precession akin to Rytov's rotation~\cite{rytov_sur_1938}.

The second contribution
\begin{equation}\label{eq:d}
    \vec d=\frac{1}{2}\mathcal{N}(\uvec w\cdot\vec{\nu})\uvec w\times\uvec e
\end{equation}
stems from $\ham_{GC}$. As shown in Fig.~\ref{fig:TPvsGC}.\textbf{a} it lies in the polarization plane, that is normal to $\uvec w$ and thus to $\vec t$. Its effect is to rotate the polarization around the ray direction $\uvec w$.  Supporting this analysis, we note that Eq.~\eqref{eq:d} is in fact analogous to the expression expected for Faraday rotation~\cite{ruiz_extending_2017}, but with the vorticity $\vec \nu$ acting here as the magnetic field. This result in turn confirms the analogy between Faraday rotation and polarization drag, extending it beyond the particular case of an aligned rotator ($\vec \nu\parallel\vec k$)~\cite{Nienhuis1992}. 

These two contributions are from symmetries completely characterized through two angles shown in Fig.~\ref{fig:TPvsGC}.\textbf{a}: the angle $\theta\in[0,\pi/2]$ between the ray direction $\uvec w$ and the local vorticity $\bm{\nu}$, and the angle $\varphi\in[0,\pi/2]$ between the projection of $\bm{\nu}$ in the plane normal to $\uvec w$ and $\uvec e$. Specifically, one finds $|\vec d|\propto\cos{\theta}$ while $|\vec t|\propto\sin{\theta}\sin{\varphi}$. The drag term $\vec d$, plotted in Fig.~\ref{fig:TPvsGC}.\textbf{b}, is indeed maximum when the ray is aligned with the vorticity ($\theta=0$), but vanishes when the vorticity is in the polarization plane ($\theta=\pi/2$), confirming the analogy with Faraday rotation. In contrast, the parallel transport plotted in Fig.~\ref{fig:TPvsGC}.\textbf{c} is zero in the aligned case. This is because, as shown in Eq.~\eqref{eq:ray_equation}, one needs a finite angle between $\uvec w$ and $\uvec \nu$ to bend the ray. The parallel transport also differs from $\vec d$ in that it depends on $\varphi$. This manifests that parallel transport depends on the projection of the polarization on the ray normal vector $d\uvec w/ds$, which from Eq.~\eqref{eq:ray_equation} is along $\uvec \nu\times\uvec w$.

Putting these pieces together, the polarization evolution along the ray writes
\begin{equation}\label{eq:unit_pola_evol}
    \fracd{\uvec e}{s}=\mathcal{N}\left(\vec\nu-\frac12(\uvec w\cdot\vec{\nu})\uvec w\right)\times\uvec e.
\end{equation}
This result confirms that a flow with non-zero vorticity induces a rotation of the polarization of the wave~\cite{rozanov_firstorder_2006}. This is illustrated in Fig.~\ref{fig:general_motion}.\textbf{a}. Note importantly that only the direction of the polarization is altered, not the polarization state. A linear polarization thus remains linear, and is simply rotated in space during propagation.

%\begin{equation}\label{eq:t}
%    t=|\bm{\nu}|\mathcal{N}\left|\left(\uvec\nu-(\uvec k\cdot\uvec{\nu})\uvec k\right)\times\uvec e\right|.
%\end{equation}

\subsection{Rotating medium}

Having shown that spin-orbit interactions and from there polarization effects arise from the vorticity of the flow, it is informative to examine more closely the canonical case of a rigid-body rotating medium, that is a constant angular velocity $\vec\Omega$. In this simple case, the vorticity $\bm{\nu }=2\vec\Omega/c$ is uniform, and the polarization evolution equation is obtained immediately from Eq.~\eqref{eq:unit_pola_evol}.

\begin{figure}
    \centering
    \includegraphics[width=\linewidth]{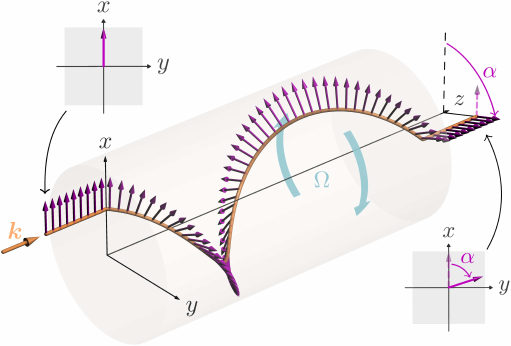}
    \caption{Helical ray and polarization rotation in the canonical example of an aligned rotator $(\vec k\parallel\vec\Omega)$. Since polarization drag supplements parallel transport in a moving medium, the ray and the polarization are rotated by the same angle $\alpha$, in contrast to the weaker polarization rotation known in static gradient index media for this same helical ray~\cite{bliokh_geometrodynamics_2009}.}
    %\caption{A ray propagating in a rotating isotropic non-dispersive medium with an helical trajectory. The initial vertical linear polarization of the wave is rotated about the rotation axis of the same angle $\theta$ as the trajectory.}
    \label{fig:rotating_aligned_case}
\end{figure}

A simple yet historically important and insightful configuration in this case is that shown in Fig.~\ref{fig:rotating_aligned_case} in which the wavevector is aligned with the rotation axis $(\vec k\parallel\vec\Omega)$. This is indeed the configuration for which Thomson~\cite{Thomson1885} and Fermi~\cite{Fermi1923} predicted polarization drag, %, as later demonstrated by Jones~\cite{jones_rotary_1976}, 
and for which a rotation of the wave's transverse structure, called image rotation~\cite{padgett_polarization_2006}, was later demonstrated~\cite{Franke-Arnold2011}. Interestingly, both polarization and image rotations are by the same angle. This dual effect of rotation has since then been explained as angular momentum coupling with, respectively, the spin and orbital angular momentum components of the wave~\cite{Goette2007}.

From the ray equation~\eqref{eq:ray_equation}, we confirm that the trajectory is then a helix wound around the rotation axis as illustrated in 
Fig.~\ref{fig:rotating_aligned_case}, with a rate of rotation of the ray consistent with image rotation. However, the Rytov's rotation expected from parallel transport along this helix is much smaller than the ray rotation. This is supported by the result that a ray following the same helix but produced in a static gradient-index medium only experiences a small change in polarization after completing a revolution along the helix~\cite{bliokh_geometrodynamics_2009}. The answer to this apparent mismatch is to be found in $\ham_{GC}$. While, because $\vec k$ is conserved when $\vec k\parallel\vec\Omega$, $\dot{\uvec w}$ and from there the parallel transport contribution are negligible to leading order in this aligned configuration, the drag term is on the other hand maximum. This corresponds to the lower right corner in Fig.~\ref{fig:TPvsGC}.\textbf{b}, where $\vec d=\mathcal{N}\vec{\Omega}\times\uvec e/c$. This drag term leads to a polarization rotation around $\vec \Omega$ by an angle $\mathcal{N}\Omega/c$ per unit length along $\vec k$, consistent with Fermi's results~\cite{Fermi1923,Thomson1885} and subsequent observations by Jones~\cite{jones_rotary_1976}. Our results thus expose the origin of the observed consistency of polarization drag and image rotation, and incidentally of the apparently supersized polarization rotation: it stems from first-order gradient corrections to the Hamiltonian induced by the flow vorticity. More generally, if the wavector is misaligned with $\vec{\Omega}$, the parallel transport contribution will add up to the drag, possibly leading to larger effects and creating new opportunities to manipulate light~\cite{gueroult_enhanced_2020}.

%As shown in Fig.~\ref{fig:rotating_aligned_case}, we find that the trajectory is then a helix wound around the rotation axis, consistent with the azimuthal drag known to lead to the rotation of the wave's transverse structure referred to as image rotation~\cite{padgett_polarization_2006, Franke-Arnold2011}. Interestingly, we note that this same trajectory can be obtained in a static gradient-index medium~\cite{bliokh_geometrodynamics_2009}. For the latter, polarization rotation stems entirely from the spin redirection and parallel transport, leading typically to a small change after one revolution along the helix. For the rotating homogeneous medium, because $\vec k$ is conserved when $\vec k\parallel\vec\Omega$, we verify that $\dot{\vec w}$ and from there the parallel transport contribution are in fact negligible to first order. On the other hand, the drag term is maximal and writes $\vec d=\mathcal{N}\vec{\Omega}\times\uvec e$, so that the polarization rotates around $\uvec \Omega$. The drag contribution then leads to a polarization rotation by an angle $\mathcal{N}\Omega/c$ per unit length along $\vec k$, which is the polarization drag angle predicted by Thomson~\cite{Thomson1885} and Fermi~\cite{Fermi1923}, and later demonstrated by Jones~\cite{jones_rotary_1976}. This rate of polarization rotation is also precisely the rate of rotation of the ray trajectory. Our results show that this equality, which has been explained as angular momentum coupling with the spin and orbital angular momentum components of the wave, stems from  

\section{Discussion}

We have shown that the non-uniform motion of a medium is the source of non-trivial spin-orbit interactions (SOIs) of light propagating through it. Specifically, we demonstrated that SOIs in moving media are akin to the spin-redirection (Rytov-Vladimirskii-Berry) phase that is well-known in static inhomogeneous media, if one replaces the wavevector and the refractive index gradient by respectively the ray direction and the local vorticity. We then exposed how this Berry phase, together with a supplemental slow phase arising from spatial derivatives of the medium's velocity field, each contributes to a change in physical polarization along the ray.

In eliciting that SOIs can be controlled by tailoring the flow of a moving medium, our results open the door for new strategies to design SOIs augmented light manipulation platforms, relying on motion instead (or in complement of) the sub-wavelength structures enabled by metamaterials. These results incidentally suggest that SOIs could also be found in time-modulated metamaterials mimicking moving matter~\cite{Bahrami2023,Caloz2023}. Properly accounting for the associated effects on polarization could also prove essential to properly interpret light signal emitted from remote sources and propagating through a moving screen, such as the polarization of light emitted by astrophysical objects~\cite{gueroult_determining_2019}. Furthermore, since geometric Berry phases have been shown to describe the parallel transport not only of the polarization but also of the beam structure~\cite{Bliokh2006}, our demonstration that non-planar rays induced by light drag lead to a non-trivial Berry phase further provides the theoretical framework to similarly study orbit-orbit coupling~\cite{bliokh_spinorbit_2015} in moving media. This includes the intrinsic orbital angular momentum Hall effect in moving media, where beam-components with a different topological charge follow different trajectories (eventually leading to beam splitting)~\cite{Bliokh2006}, as elicited in the particular case of a rotating electromagnetically induced transparency medium~\cite{Zhao2019}. Finally, in making use of Wigner-Weyl transforms, our first-order eikonal model for light in moving media can readily be extended to more complex wave equations characterizing more complex media, paving the way for the discovery of new class of effects in moving media.

\section*{Acknowledgments}

This work is supported by the Agence Nationale de la Recherche (ANR) through the WaRP project (ANR-21-CE30-0002). This work has been carried out within the framework of the EUROfusion Consortium, via the Euratom Research and Training Programme (Grant Agreement No 101052200 – EUROfusion). Views and opinions expressed are however those of the authors only and do not necessarily reflect those of the European Union or the European Commission. 

The authors thank J. Langlois for constructive discussions.

\appendix

\section{Wave equation for a moving isotropic non-dispersive medium}\label{app:A}

Consider a isotropic non-dispersive medium. Assuming that it is homogeneous and stationary, let us write $n'$ its refraction index, which is constant over space and time. The constitutive relations describing this medium's response in the reference frame $\Sigma'$ in which it is at rest then simply write
\begin{subequations}\label{eq:constitutive_relations_rest-frame}
    \begin{align}
        \vec D'&=\epsilon_0 n'^2\vec E'\\
        \vec B'&=\mu_0\vec H',
    \end{align}
\end{subequations}
with primes denoting quantities in $\Sigma'$. 

If this medium is now moving with respect to an observer at rest in the laboratory frame $\Sigma$, its response as seen from $\Sigma$ is modified. As originally derived by Minkowski~\cite{Minkowski1908}, the constitutive relations of the moving medium in $\Sigma$ are then obtained by performing a local and instantaneous Lorentz transformation of the primed fields in Eqs.~\eqref{eq:constitutive_relations_rest-frame}. Considering the motion to be stationary, $\vec\beta(\vec x)=\vec v(\vec x)/c$, they classically read to first order in $\beta$  
\begin{subequations}\label{eq:constitutive_relations_lab-frame}
    \begin{align}
        \vec D&=\epsilon_0n'^2\vec E+(n'^2-1)\vec\beta\times\vec H/c\\
        \vec B&=\mu_0\vec H-(n'^2-1)\vec\beta\times\vec E/c.
    \end{align}
\end{subequations}
Plugging Eqs.~\eqref{eq:constitutive_relations_lab-frame} into Maxwell equations yields a pair of coupled equations for $\vec E$ and $\vec H$ only
\begin{subequations}
    \begin{align}\label{eq:coupled_eq_curlE}
        \vec\nabla\times\vec E&=-\mu_0\partial_t\vec H+(n'^2-1)\vec\beta\times\partial_t\vec E/c\\\label{eq:coupled_eq_curlH}
        \vec\nabla\times\vec H&=\epsilon_0n'^2\partial_t\vec E+(n'^2-1)\vec\beta\times\partial_t\vec H/c.
    \end{align}
\end{subequations}
Finally, taking the curl of Eq.~\eqref{eq:coupled_eq_curlE} and using Eqs.~\eqref{eq:coupled_eq_curlH} and~\eqref{eq:coupled_eq_curlE} to eliminate the terms $\vec\nabla\times\vec H$ and $\partial_t\vec H$, a unique wave equation for the electric field $\vec E$ is obtained. Keeping only terms up to first order in $\beta$, it writes
\begin{multline}\label{eq:wave_equation_methods}
        \vec\nabla\times\vec\nabla\times\vec E+n'^2\partial_{ct}^2\vec E= 
        \\(n'^2-1)\left[\vec\beta\times\vec\nabla\times\partial_{ct}\vec E+\vec\nabla\times\left(\vec\beta\times\partial_{ct}\vec E\right)\right],
\end{multline}
where we introduced the notation $\partial_{ct}=\partial_t/c$. This is Eq.~\eqref{Eq:wave_eq}.

\section{Dispersion symbol and eigenmodes of the moving medium}\label{app:B}

Using operator formalism, and defining $\widetilde{\vec k}=i\vec\nabla$ and $\widetilde{\omega}=-i\partial_t$, the wave equation Eq.~\eqref{eq:wave_equation_methods} can be rewritten $\tensop D\vec E=0$ with the wave operator
\begin{multline}
\label{Eq:wave_operator}
    \tensop D(\widetilde {\vec x},\widetilde {\vec k},\widetilde\omega)=\left[n'^2\widetilde{\omega}^2/c^2-\widetilde{\vec k}\cdot\widetilde{\vec k}\right]\widetilde{\eye}+\widetilde{\vec k}\otimes\widetilde{\vec k}\\
    -(n'^2-1)\left[\widetilde{\omega}\tensor\beta(\widetilde{\vec x})\widetilde{\tensor k}+\widetilde{\tensor k}\tensor\beta(\widetilde{\vec x})\widetilde{\omega}\right]/c.
\end{multline}
Here and henceforth an underlined variable indicates a matrix and an underlined vector quantity refers to the cross-product antisymmetric matrix associated to it, i.e. $\tensor k\vec a=\vec k\times\vec a$. Making use of symbolic calculus, we then define the symbol $\tensor D$ of this operator $\tensop D$, which is formally the Wigner transform of the wave operator, and is called the dispersion symbol~\cite{tracy_ray_2014,dodin_quasioptical_2019,venaille_ray_2023}. For the relatively simple operator $\tensop D$ in Eq.~\eqref{Eq:wave_operator}, an expression for its symbol $\tensor D$ can in fact be obtained simply from elementary Wigner transforms (see, e.g., supplementary materials in Ref.~\cite{dodin_quasioptical_2019}). 

Mirroring the geometrical optics (GO) expansion in the small parameter $\varepsilon$, which we recall here characterizes the velocity field non-uniformity $\varepsilon\doteq\lambda\|\vec\nabla\vec\beta\|_\infty/\beta$, we decompose the symbol $\tensor D$ into its zeroth- and first-order parts~\cite{dodin_quasioptical_2019,onuki_quasi-local_2020,venaille_ray_2023}, denoted respectively $\tensor D_0$ and $\tensor D_1$. After some algebra we eventually obtain
\begin{multline}
    \label{eq:dispersion_matrix}
    \tensor D_0(\vec x,\vec k,\omega)=\left[n'^2\frac{\omega^2}{c^2}-\vec k^2\right]\eye+\vec k\vec k^T\\+(n'^2-1)\frac{\omega}{c}\left[\tensor\beta(\vec x)\tensor k+\tensor k\tensor\beta(\vec x)\right]
\end{multline}
and
\begin{equation}\label{eq:dispersion_symbol_first-order_part}
    \tensor D_1(\vec x,\vec k,\omega)=\frac i2(n'^2-1)\frac{\omega}{c}\tensor{\nu}(\vec x).
\end{equation}

Examining the dispersion matrix $\tensor D_0$ closer, we find that it is identical to first order in $\beta$ to
\begin{align}\label{eq:modified_dispersion_matrix}
    \tensor D_{0_w}=\ham_0\eye+\vec w\vec w^T,
\end{align}
with 
\begin{multline}\label{eq:dispersion_function}
    \ham_0(\vec x,\vec k,\omega)=\frac{\omega^2}{c^2}-\vec k^2\\+(n'^2-1)\frac{\omega}{c}\left[\frac{\omega}{c}-2\vec\beta(\vec x)\cdot\vec k\right]
\end{multline}
and 
\begin{align}\label{Eq:def_omega}
    \vec w(\vec x,\vec k,\omega)=\vec k+(n'^2-1)\frac{\omega}{c}\vec \beta(\vec x).
\end{align}
We argue here that $\tensor D_{0_w}$ is in fact the physically correct dispersion matrix since it ensures that the Poynting vector is aligned with the group velocity, as expected for a moving non-dispersive medium~\cite{ko1977energy}. Indeed, in contrast to $\tensor D_0$, it guarantees that the electric field is normal to $\vec w$ which will be shown to be aligned with the group velocity. Another argument is that a form similar to Eq.~\eqref{eq:modified_dispersion_matrix} is obtained for the electromagnetic four-potential under a full relativistic treatment, suggesting that the difference between $\tensor D_{0}$ and $\tensor D_{0_w}$ is an artifact of working with a wave equation valid up to first order only. Supporting further this last argument, one verifies that $\ham_0$ is indeed the first-order expansion of the dispersion function obtained from Gordon’s optical metric~\cite{gordon_zur_1923}.

Eq.~\eqref{eq:modified_dispersion_matrix} shows immediately the existence of two degenerate transverse modes (i.~e. normal to $\vec w$) with the common dispersion function $\ham_0$ from Eq.~\eqref{eq:dispersion_function}. Here we choose to work with circularly polarized modes and use $+$ and $-$ subscripts to refer to respectively the right- and the left-circular polarizations. The two unit circularly-polarized eigenvectors corresponding to the eigenvalue $\ham_0$ are then $\uvec \eta_{\pm}$ with
\begin{equation}\label{Eq:eigenvectors}
    \vec\eta_\pm(\vec x,\vec k,\omega)=\left(\eye-\uvec w\uvec w^T\pm i\utensor w\right)\uvec u,
\end{equation}
where an overhat refers to a unit length vector, $\uvec a = \vec a/|\vec a|$, and the vector field $\uvec u=\uvec u(\vec x,\vec k,\omega)$ is a gauge vector field which can be chosen arbitrarily without affecting the physics, as long as it is not aligned with $\uvec w$. This gauge choice simply manifests the fact that, as usual, the eigenvectors are known up to an arbitrary phase ($U(1)$ gauge)~\cite{littlejohn_geometric_1991}. Here we choose for simplicity a constant vector $\uvec u$ as the gauge vector field.

\section{Classical geometrical optics ray equations}\label{app:C}

With the dispersion function $\ham_0$ for the modes in hand, we can compute the ray equations which describe in the geometrical optics approximation the trajectory of a wave packet in phase space $(\vec x,\vec k)$. Mathematically, the ray equations are analogous to Hamilton's equations, and can be written
\begin{subequations}\label{eq:ray_equations_general}
    \begin{align}
        \fracd{\vec x}{s}&=-\vec\nabla_{\vec k}\ham_0/|\vec\nabla_{\vec k}\ham_0|\\
        \fracd{\vec k}{s}&=\vec\nabla_{\vec x}\ham_0/|\vec\nabla_{\vec k}\ham_0|
    \end{align}
\end{subequations}
where the ray parameter $s$ corresponds here to the physical length of the ray~\cite{bliokh_geometrodynamics_2009}. Using $\ham_0$ from Eq.~\eqref{eq:dispersion_function} and defining $\mathcal N=\left(n'-1/n'\right)$, the ray equations in a moving isotropic non-dispersive medium with an arbitrary velocity field then write, to first order in $\beta$, 
\begin{subequations}\label{eq:ray_equations}
    \begin{align}\label{eq:ray_equation_x}
        \fracd{\vec x}{s}&=\uvec k+\mathcal N\vec\beta\\
        \fracd{\vec k}{s}&=\mathcal N\left[(\vec\nabla\times\vec\beta)\times\vec k-(\vec k\cdot\vec\nabla)\vec\beta\right]\label{eq:ray_equation_k},
    \end{align}
\end{subequations}
consistent with Leonhardt's~\cite{leonhardt_optics_1999}. We verify that the right-hand side in Eq.~\eqref{eq:ray_equation_x} is equal to first order in $\beta$ to $\uvec w=\vec w/|\vec w|$ as computed from Eq.~\eqref{Eq:def_omega}. This supports our definition $\uvec w\doteq d\vec x/ds$. Besides, since by construction $d\vec x/ds$ is aligned with the group velocity, it also confirms as postulated earlier that $\vec w$ is aligned with the group velocity. Differentiating Eq.~\eqref{eq:ray_equation_x} with respect to $s$, eliminating $d\vec k/ds$ using Eq.~\eqref{eq:ray_equation_k}, and finally rewriting $\vec k$ in terms of $d\vec x/ds$ using Eq.~\eqref{eq:ray_equation_x}, yields an equation for the position $\vec x$ only, which is Eq.~\eqref{eq:ray_equation}.

Note for completeness that there is a third ray equation on the time coordinate $t$ supplementing Eqs.~\eqref{eq:ray_equations_general}, that is $dt/ds=\partial_\omega\ham_0/|\vec\nabla_{\vec k}\ham_0|$. It is required if one wants to compute the group velocity $\vec v_g=d\vec x/dt=(d\vec x/ds)(ds/dt)=-(\partial_\omega\ham_0)^{-1}\vec\nabla_{\vec k}\ham_0$.

\section{First-order Hamiltonian}\label{app:D}

Going back to quasi-plane waves, the evolution of the envelope, and from there of the physical polarization, is governed by the first-order part $\tensor\ham_1$ of the Hamiltonian describing the wave to first order in $\epsilon$. Relying on Wigner-Weyl transforms, Ruiz and Dodin showed~\cite{ruiz_extending_2017,dodin_quasioptical_2019} that, for any linear wave equation, $\tensor\ham_1$ can be written as the sum of two terms, denoted here $\tensor\ham_{SOI}$ and $\tensor\ham_{GC}$, expressing these terms through the zeroth and first-order parts of the dispersion symbol Eqs.~\eqref{eq:modified_dispersion_matrix} and \eqref{eq:dispersion_symbol_first-order_part} and a polarization matrix gathering the mode eigenvectors $\tensor\Xi=\begin{pmatrix}\uvec \eta_+&\uvec\eta_-\end{pmatrix}$. Note that here we write $\tensor\ham_1$ the negative first-order correction of the Hamiltonian $\tensor\ham=\ham_0\eye-\tensor\ham_1$, consistent with Ref.~\cite{dodin_quasioptical_2019}.

Specifically, they showed that the spin-orbit contribution $\tensor\ham_{SOI}$ can, for a stationary medium, be written~\cite{dodin_quasioptical_2019}
\begin{equation}\label{eq:general_H_SOI}
    \tensor\ham_{SOI}=-\dot{x}_i\tensor\cala^{(x)}_i-\dot{k}_i\tensor\cala^{(k)}_i
\end{equation}
where the space and momentum Berry connections are matrices respectively given by
\begin{subequations}
    \begin{align}       \tensor\cala^{(x)}_i&=\left(\tensor\Xi^\dag\partial_{x_i}\tensor\Xi\right)_A\\
\tensor\cala^{(k)}_i&=\left(\tensor\Xi^\dag\partial_{k_i}\tensor\Xi\right)_A.
    \end{align}
\end{subequations}
Here repeated indices are implicitly summed, and the notation $(.)_A$ refers to the antisymmetric part of a matrix. Computing these matrix connections for the circularly polarized eigenvectors given in Eq.~\eqref{Eq:eigenvectors}, we find that they are diagonal. The diagonal components can then more classically be written as the vector connections
\begin{equation}
    \vec\cala^{(x)}_\pm=\pm \mathcal{N}\frac{\uvec k\cdot\uvec u}{1-(\uvec k\cdot\uvec u)^2}(\vec\nabla\vec\beta)(\uvec k\times\uvec u)
\end{equation}
and
\begin{equation}
    \vec\cala^{(k)}_\pm=\pm\frac{\uvec k\cdot\uvec u}{1-(\uvec k\cdot\uvec u)^2}(\uvec k\times\uvec u).
\end{equation}
Plugging these results together with the ray equations~\eqref{eq:ray_equations} in Eq.~\eqref{eq:general_H_SOI} finally yields 
\begin{equation}
    \ham_{SOI\pm}=\pm\mathcal N \frac{\uvec k\cdot\uvec u}{1-(\uvec k\cdot\uvec u)^2}\left(\vec\nu\cdot\uvec u-(\vec\nu\cdot\uvec k)(\uvec k\cdot\uvec u)\right).
\end{equation}
This is the spin-orbit contribution to the first-order part of the Hamiltonian in a moving medium, which is clearly shown here to depend on the local vorticity $\vec \nu$. Note that while $\ham_{SOI\pm}$ depends on the gauge $\uvec u$, this gauge dependence will vanish as expected in equations on physical observables, as we will show for the polarization evolution equation. Finally, remarking that one may equivalently write $\ham_{SOI\pm}=i\uvec \eta_\pm^\dag d\uvec\eta_\pm/ds$~\cite{littlejohn_geometric_1991}, and because $\uvec \eta$ only depends on $\uvec w$, we note that an alternative form is
\begin{equation}
    \ham_{SOI\pm}=i\uvec \eta_\pm^\dag \dot{\hat w}_j\frac{\partial\uvec\eta_\pm}{\partial\hat w_j}.
\end{equation}
This rewriting then suggests to define $\vec\cala^{(\hat w)}_\pm $ the Berry connection associated with the ray tangent $\uvec w$, so that
\begin{equation}
    \ham_{SOI\pm}=-\dot{\uvec w}\cdot\vec\cala^{(\hat w)}_\pm.
\end{equation}
After some algebra we find
\begin{equation}
    \vec\cala^{(\hat w)}_\pm=\pm\frac{\uvec w\cdot\uvec u}{1-(\uvec w\cdot\uvec u)^2}(\uvec w\times\uvec u),
\end{equation}
as given in Eq.~\eqref{Eq:Connection_w}.

Moving on to the the gradient correction contribution, it was shown to write for a stationary medium~\cite{dodin_quasioptical_2019,ruiz_extending_2017}.
\begin{equation}\label{Eq:H_gc_general}
\tensor\ham_{GC}=\left(\partial_{x_i}\tensor\Xi^\dag\left(\tensor D_0-\ham_0\eye\right)\partial_{k_i}\tensor\Xi\right)_A-\tensor\Xi^\dag\tensor D_1\tensor\Xi
\end{equation}
After some algebra we verify that this matrix is also diagonal with diagonal components
\begin{equation}
    \ham_{GC\pm}=\mp\frac{1}{2}\mathcal N\uvec w\cdot\vec\nu,
\end{equation}
that is Eq.~\eqref{eq:H_GC}. Note that we used here $\tensor D_{0_w}$ in lieu of $\tensor D_0$ in Eq.~\eqref{Eq:H_gc_general} for the reasons outlined above, but with no effect since we work here to first order in $\beta$.

\section{Evolution equation for the polarization}\label{app:E}

Putting these pieces together, we can now examine the effect of first-order corrections to the Hamiltonian on the wave's polarization. For this we introduce 
\begin{equation}
    \uvec e=\left(\vec E_++\vec E_-\right)e^{-i\Phi}/|\vec E|.
\end{equation}
the unit polarization vector, which captures the state and the direction of the wave's polarization. For quasi-plane waves of the form $\vec E_\pm=|a_\pm|e^{i\psi_\pm}\uvec\eta_\pm e^{i\Phi}$, this rewrites
\begin{equation}\label{eq:unit_pola_explicit}
    \uvec e=\varkappa_+e^{i\psi_+}\uvec\eta_++\varkappa_-e^{i\psi_-}\uvec\eta_-
\end{equation}
with $\varkappa_\pm=|a_\pm|/|\vec E|$ the normalized amplitude of the modes. Differentiating Eq.~\eqref{eq:unit_pola_explicit} with respect to the ray parameter $s$ then yields
\begin{equation}\label{eq:general_pola_evol_eq}
    \fracd{\uvec e}{s}=\sum_{m=\pm} \left(\fracd{\varkappa_m}{s}\uvec\eta_m+\varkappa_m\fracd{\uvec\eta_m}{s}+i\varkappa_m\fracd{\psi_m}{s}\uvec\eta_m\right)e^{i\psi_m}.
\end{equation}

The middle term on the right-hand side is known through the eigenvectors $\uvec \eta_\pm$ in Eq.~\eqref{Eq:eigenvectors}, but we still need $d\varkappa_\pm/ds$ and $d\psi_\pm/ds$. For that we use 
here that the evolution of the unit complex two-component Jones vector $\uvec \xi=\begin{pmatrix}\varkappa_+e^{i\psi_+}&\varkappa_-e^{i\psi_-}\end{pmatrix}^T$ is governed by~\cite{Bliokh2007,Ma2016,ruiz_extending_2017}  
\begin{equation}\label{eq:Jones_vector_evol_eq}
    \fracd{\uvec \xi}{s}=i\tensor\ham_1\uvec \xi
\end{equation}
Because $\tensor\ham_1$ is diagonal in the case of interest here, Eq.~\eqref{eq:Jones_vector_evol_eq} decouples into two independent scalar equations $d(\varkappa_{\pm}e^{i\psi_{\pm}})/ds=i\ham_{1\pm}\varkappa_{\pm}e^{i\psi_{\pm}}$. Moreover, since $\ham_{1\pm}$ are here real scalars, we get $d\varkappa_\pm/ds=0$ and hence
\begin{equation}
    \fracd{\psi_\pm}{s}=\ham_{1\pm}=\ham_{SOI\pm}+\ham_{GC\pm}.
\end{equation}
The evolution of the slow phases $\psi_\pm$ is thus governed by the first-order Hamiltonian corrections. 

Using these results, we find that the polarization evolution equation in a moving media can be written to lowest order in $\varepsilon$ and $\beta$
\begin{equation}
    \fracd{\uvec e}{s}=\vec t+\vec d
\end{equation}
with 
\begin{subequations}
    \begin{align}
        \vec t & =\sum_{m=\pm} \varkappa_m\left(\fracd{\uvec\eta_m}{s}+i\ham_{SOI_m}\uvec\eta_m\right)\\\label{Eq:t_appendix}
         & =\mathcal N\left(\vec\nu-(\uvec w\cdot\vec\nu)\uvec w\right)\times\uvec e\
\end{align}
\end{subequations}

and 
\begin{subequations}
    \begin{align}\label{eq:d_appendix}
        \vec d&=i\sum_{m=\pm}\varkappa_m\ham_{GC_m}\uvec\eta_m\\
         & = \frac12\mathcal N(\uvec w\cdot\vec\nu)\uvec w\times\uvec e,
    \end{align}
\end{subequations}
which are respectively Eqs.~\eqref{eq:t} and \eqref{eq:d}. As expected, we verify here that both $\vec t$ and $\vec d$ are gauge-independent. One further verifies that the expression on the right-hand side in Eq.~\eqref{Eq:t_appendix} is equal to the well-known parallel transport variation $-(\uvec e\cdot\dot{\uvec w})\uvec w$~\cite{bliokh_geometrodynamics_2009}, supporting the physical interpretation of this SOI contribution.

%Using these results, Eq.~\eqref{eq:general_pola_evol_eq} rewrites
%\begin{equation}
%    \fracd{\uvec e}{s}=\sum_{m=\pm} a_m\left(\fracd{\uvec\eta_m}{s}+i\ham_{SOI_m}\uvec\eta_m+i\ham_{GC_m}\uvec\eta_m\right).
%\end{equation}

% \bibliography{references,refs}% Produces the bibliography via BibTeX.


\end{document}